\documentstyle[prl,aps,eqsecnum,epsfig,multicol]{revtex}

\begin{document}
\draft

\title{Measurement of the $\Lambda$ and $\bar{\Lambda}$ particles \\
        in  Au+Au Collisions at $\sqrt{s_{_{NN}}}~=~130$~GeV}

\author{
K.~Adcox,$^{40}$
S.{\,}S.~Adler,$^{3}$
N.{\,}N.~Ajitanand,$^{27}$
Y.~Akiba,$^{14}$
J.~Alexander,$^{27}$
L.~Aphecetche,$^{34}$
Y.~Arai,$^{14}$
S.{\,}
% ...
\\(PHENIX Collaboration)\\
}
\address{
% ...
$^{3}$Brookhaven National Laboratory, Upton, NY 11973-5000, USA\\
% ...
$^{14}$KEK, High Energy Accelerator Research Organization, Tsukuba-shi, Ibaraki-ken 305-0801, Japan\\
% ...
$^{27}$Chemistry Department, State University of New York - Stony Brook, Stony Brook, NY 11794, USA\\
% ...
$^{34}$SUBATECH (Ecole des Mines de Nantes, IN2P3/CNRS, Universite de Nantes) BP 20722 - 44307, Nantes-cedex 3, France\\
% ...
$^{40}$Vanderbilt University, Nashville, TN 37235, USA\\
% ...
}

\date{\today}

\maketitle

\begin{abstract}
We present results on the measurement of $\Lambda$ and $\bar{\Lambda}$ 
production 
in  Au+Au collisions at $\sqrt{s_{_{NN}}}~=~130$~GeV with the PHENIX 
detector at RHIC. The transverse momentum spectra were measured for
minimum--bias and for the 5$\%$ most central events.
The $\bar{\Lambda}/\Lambda$ ratios are constant
as a function of $p_T$ and the number of participants.
The measured net $\Lambda$ density is significantly
larger than predicted by models based on hadronic strings (e.g. HIJING)
but in approximate agreement with models which include the gluon junction 
mechanism.
\end{abstract}

\begin{multicols}{2}
\narrowtext

In this paper we report on the measurement by the PHENIX experiment
at the Relativistic Heavy Ion Collider (RHIC)
of the production of $\Lambda$ and $\bar{\Lambda}$ particles and
the ratio $\bar{\Lambda}/\Lambda$
as a function of transverse momentum $p_T$ and centrality in
Au+Au collisions at $\sqrt{s_{NN}}$~=~130~GeV. The production of
strange baryons and of strange particles in general has been
extensively studied in heavy ion collisions at the
Alternating Gradient Synchrotron (AGS) and at the
Super Proton Synchrotron (SPS) \cite{sqm-2000}
to investigate the flavor composition of nuclear matter at
high density and temperature.
Furthermore, antibaryon--to--baryon ratios,
or alternatively, a net baryon number such as ($\Lambda-\bar{\Lambda}$)
at midrapidity provide insight into the baryon transport mechanism
in these collisions \cite{kopeliovich,kharzeev,vance,gyalassy}.
The systematic study of baryon stopping
(transport of baryon number in rapidity space) and hyperon production
in proton--proton, proton--nucleus and nucleus--nucleus collisions
has been done over the past decade at the AGS~\cite{ahmed,filimonov}
and CERN SPS~\cite{na35,na44,na49}.
The results have shown a high degree of baryon stopping and
enhanced hyperon production in heavy nucleus--nucleus collisions.
Clearly the measurement of these quantities at RHIC, at the highest energies
so far available in the laboratory, is of great importance for our understanding
of the sources of these processes.

The results reported here were obtained using the west arm of the PHENIX
spectrometer \cite{phenix_1} which covers an angular range of
$\Delta\phi$~=~$\pi$/4 (during its first year of running) and a pseudorapidity range of
$|\eta|$~$<$~0.35. The detectors used were the drift chamber (DC),
a set of multiwire proportional chambers with pixel--pad readout (PC1) \cite{pc}, 
and a lead--scintillator electromagnetic
calorimeter (EMCal) \cite{emcal}. 
Signals from two sets of beam--beam counters (BBC) and two
zero--degree calorimeters (ZDC) provided a trigger sensitive to 92
$\pm$ 4$\%$ of the 6.8 barn total Au+Au cross section \cite{ppg001}. 
The centrality selection was done using the correlation between the analog responses of the ZDC and BBC \cite{ppg001}.

The present analysis is based on
1.3~M minimum--bias events with a vertex position of $|z|$~$<$~20~cm.
To reconstruct the $\Lambda$ and $\bar{\Lambda}$ particles,
their weak decays $\Lambda\rightarrow p\pi^-$
and $\bar{\Lambda}\rightarrow \bar{p}\pi^+$ are used.
The tracks of the charged particles from 
the decay of $\Lambda$ and $\bar{\Lambda}$ are reconstructed using DC and PC1, 
and their momentum is determined by the DC with a resolution of 
$\delta p/p~\simeq~0.6\%~\oplus~3.6\%~p$~(GeV/c).
Although these tracks do not point back
to the primary vertex position their momentum is calculated assuming that they come from
the interaction point, hence there is in general a shift in the
momentum of the reconstructed $\Lambda$ and $\bar{\Lambda}$ particles. 
A Monte Carlo (MC) study shows that the difference is of the order
1--2$\%$ over the measured momentum range, within the measured
momentum resolution, and thus neglected in the present study.
The absolute momentum
scale is known to better than 2$\%$ \cite{ppg003}. In order to reduce background the 
tracks are confirmed by requiring a matching hit in the EMCal
within $\pm 3\sigma$.
For the particle identification (PID) of the daughter charged particles the
time--of--flight signal of the EMCal with a timing resolution 
of $\sim$~700~ps is used. Using the momentum measured by the DC
and the flight--time, the particle mass is calculated and 
a 2$\sigma$ momentum--dependent cut is applied to the mass--squared 
distribution to identify protons,
antiprotons and pions. An upper momentum cut of 0.6 GeV/c and 1.4
GeV/c for pions and protons respectively provides clean particle separation.
Then each proton is
combined with each pion in the same event and the invariant mass is calculated.
If $\Lambda$ or $\bar{\Lambda}$
are produced a peak appears in the mass distribution on top of a background
from random combinations of particles. In order to determine the number of
$\Lambda$'s and $\bar{\Lambda}$'s from such a distribution an estimation of the background
is essential. For this, the mass distribution with combinations of protons and pions
from different events with the same centrality class (event mixing) is used.
In order to reduce the combinatorial background the decay proton energy
 is required to be within
$E_{p}^{min}<E_{p}<E_{p}^{max}$ where $E_{p}^{min}$
and $E_{p}^{max}$ are calculated from the two--body decay kinematics in the 
$\Lambda$ center--of--mass system. A similar cut
is used for the pions. This cut gives an improvement of the signal--to--background
ratio by a factor of two and results in the final value of S/B = 1/2 for both
$\Lambda$ and $\bar{\Lambda}$. We obtain $\sim$~12000 $\Lambda$ and 
$\sim$~9000 $\bar{\Lambda}$ particles in the mass range
1.05~$<$~m$_{p\pi}~<$~1.20~GeV/c$^2$
with mass resolution $\delta m/m~\simeq~2\%$ obtained from a Gaussian fit.

\begin{figure}
\centerline{\epsfig{file=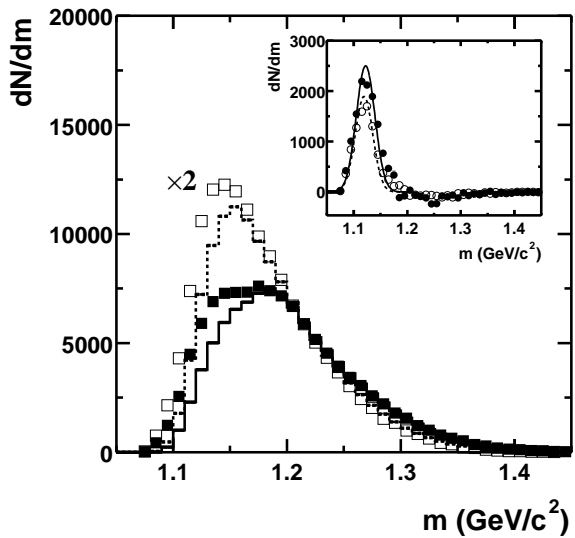,width=1.0\linewidth}}
\caption[]{Invariant mass spectra of $p\pi^-$ (solid squares)
and $\bar{p}\pi^+$ (open squares) pairs. The histograms show the background
for $p\pi^-$ (solid) and $\bar{p}\pi^+$ (dashed).
For clarity the $\bar{p}\pi^+$ data are scaled up by a factor of two.
The insert shows the
background--subtracted spectra of $\Lambda$ (solid circles) 
and $\bar{\Lambda}$ (open circles).
The lines are Gaussian fits to the
$\Lambda$ (solid line) and $\bar{\Lambda}$ (dashed line) mass spectra.}
\label{fig:lambda_cut_new}
\end{figure}

Fig.~\ref{fig:lambda_cut_new}  shows the invariant mass 
spectra for the $\Lambda\rightarrow p\pi^-$
and $\bar{\Lambda}\rightarrow \bar{p}\pi^+$.
The results represent the primary $\Lambda$ and $\bar{\Lambda}$ and contributions
from the feed--down from heavier hyperons (mainly $\Sigma^0$ and $\Xi$).
The reconstructed number of $\Lambda$ and $\bar{\Lambda}$
particles is corrected for the acceptance, pion decay--in--flight, momentum 
resolution and 
reconstruction efficiency. For this, single--particle MC events were generated
over the full azimuth $\phi$ (0~$<~\phi~<~2\pi$) and one unit of rapidity $y$ 
(--0.5~$<$~$y$~$<$~0.5). The simulated particles were passed through the entire 
PHENIX GEANT \cite{geant} simulation. The correction function is defined as the ratio 
of the input (generated) transverse momentum distribution to the $p_T$
distribution of the particles reconstructed in the spectrometer. However  this correction 
does not take into account the decreasing track reconstruction efficiency due 
to the high multiplicity environment in more central events. A well established
method to obtain this efficiency drop is the embedding procedure 
\cite{ppg003}. We use
single--particle MC tracks embedded into real events and analyze the merged events
with the same analysis code as used for the reconstruction of the data set.
The track reconstruction efficiency decreases from 90$\%$ for minimum--bias 
to 70$\%$ for central events.
The $\Sigma^0$ and $\Xi$ hyperons decay to $\Lambda$ with a branching ratio
of essentially 100$\%$.  We have verified, using HIJING, that the kinematic
properties ($p_T$ and $y$ distributions) of primary $\Lambda$ and those produced
by $\Sigma^0$ and $\Xi$ decay are the same (within a few percent).
We conclude that the correction
function determined for primary $\Lambda$ is valid for all $\Lambda$.  Since
there are no reliable data available for the yields of those hyperons at
RHIC energies, we cannot quantitatively state the contributions from those
heavier hyperons to our $\Lambda$ production.  Therefore, our data include
the primordial $\Lambda$ and $\bar{\Lambda}$, as well as the feed--down from the
heavier hyperons.

\begin{figure}
\centerline{\epsfig{file=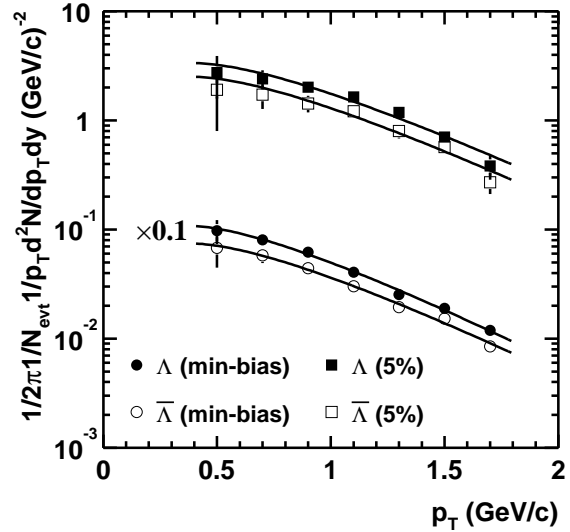,width=1.0\linewidth}}
\caption[]{Transverse momentum spectra of $\Lambda$ and $\bar{\Lambda}$ for
                 minimum--bias and for the 5$\%$ most central events.
                 For clarity of presentation 
                 the data points for minimum--bias are scaled down by a factor of ten.}
\label{fig:yields}
\end{figure}

Using the correction function determined from single--particle MC and
multiplicity--dependent track reconstruction efficiency derived from the
embedding procedure we correct the transverse
momentum distributions of $\Lambda$ and $\bar{\Lambda}$.
The invariant yields as a function of 
the transverse momentum $p_T$ for $\Lambda$ and $\bar{\Lambda}$
are shown in Fig.~\ref{fig:yields} for minimum--bias events (circles).
Over the measured transverse momentum range
(0.4~$<$~$p_T$~$<$~1.8~GeV/c) both the $\Lambda$ and $\bar{\Lambda}$
spectra can be described by a distribution of the form 
$dN/{{dp_T}^2}~\propto~p_Te^{-p_T/T}$
as shown in Fig.~\ref{fig:yields} by the solid line. 
The total yield $dN/dy$ is
obtained by integrating the functional fit from zero to infinity. In the same
Fig.~\ref{fig:yields} we show the invariant yield for the 5$\%$ most
central events (squares).
In Table~\ref{tab:table1} the total yields $dN/dy$ together with the
average transverse momentum $<\!\!p_T\!\!>$
and the slope parameters $T$
are listed for both minimum--bias and for the 5$\%$ most central events.

There are several sources which contribute to the systematic uncertainties of the
total yield $dN/dy$ and average transverse momentum $<\!\!p_T\!\!>$ determination.
One of the sources, the uncertainty in the correction function determination,
is found to be 13$\%$.
A second contribution is due to the fitting function used for the
extrapolation beyond the measured $p_T$ range: the fraction of the
extrapolated yield is 29~$\pm$~9~$\%$. There is also an additional contribution to the
systematic error which originates from the combinatorial background subtraction which is
3$\%$. Combining the above uncertainties in quadrature
gives a total systematic error in the yield of 16$\%$ (see Table~\ref{tab:table1}).

\begin{figure}
\centerline{\epsfig{file=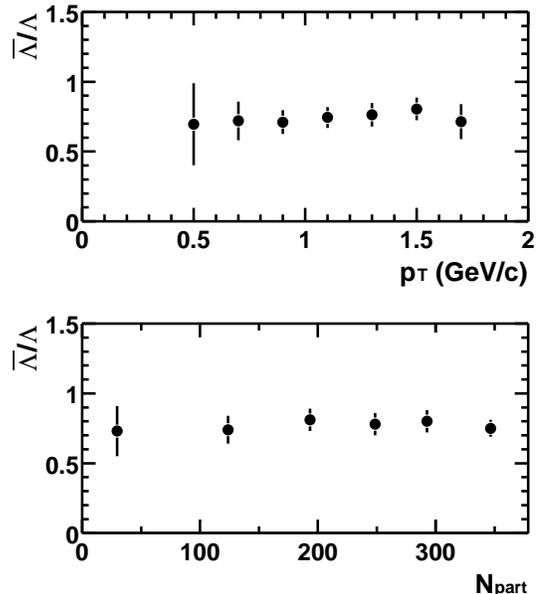,width=1.0\linewidth}}
\caption[]{The ratio $\bar{\Lambda}$$/$$\Lambda$ as a function of $p_T$ (top) and as a function of the number of participants (bottom).}
\label{fig:ratios}
\end{figure}

The $\bar{\Lambda}/\Lambda$ yield ratio determined in the mass window defined 
above versus 
$p_T$ is shown in Fig.~\ref{fig:ratios} (top panel).
The ratio is constant over the whole $p_T$ range  0.4~$<~p_T~<$~1.8~GeV/c.
There is also no significant variation of the $\bar{\Lambda}/\Lambda$ ratio
as a function of the number of participants (bottom panel) which is calculated  using a Glauber 
model together with a simulation of the ZDC and BBC responses \cite{ppg001}.
The average $\bar{\Lambda}/\Lambda$ ratio is 0.75~$\pm$~0.09 (stat).
Both the $p_T$--dependence and the integral
$\bar{\Lambda}/\Lambda$ ratio are consistent with statistical thermal
model calculations \cite{thermal} at RHIC energies.

\begin{figure}
\centerline{\epsfig{file=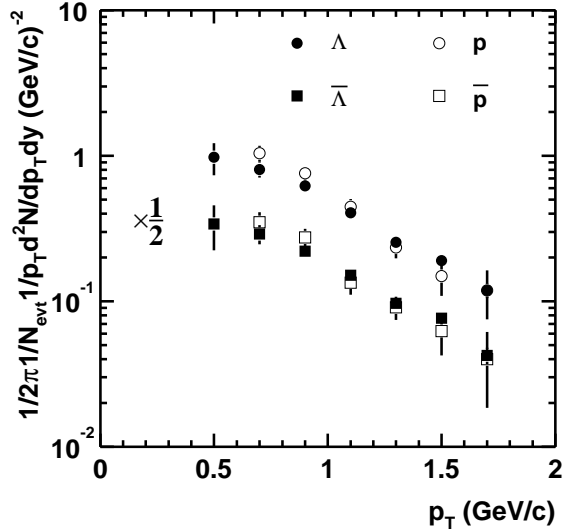,width=1.0\linewidth}}
\caption[]{Spectra of inclusive $\Lambda$ ($\bar{\Lambda}$)
and feed--down corrected protons (antiprotons) for minimum--bias events. The data points for antiprotons and $\bar{\Lambda}$ are scaled down by a factor of two.}
\label{fig:protons}
\end{figure}

The present measurement of the total yield of $\Lambda$ and $\bar{\Lambda}$
enables us to take the previously reported inclusive $p$ and $\bar{p}$ spectra \cite{ppg006},
and construct a feed--down correction for $\Lambda$ decays.
As the $\Sigma$ yield has not been measured, we do not inlcude feed--down
from $\Sigma^{\pm}$ decays, but this is expected to be $<~5\%$, based
on HIJING calculations. The feed--down corrections were done bin by bin
on the proton (antiproton) $p_T$ spectrum by the following method:

\begin{equation}
\frac{dN^{p}}{dyd{p_T}}(i) = \frac{dN^{m}}{dyd{p_T}}(i)-\sum_{j=1}^{N_{bins}} 
                                       \frac{dN^{\Lambda}}{dyd{p_T}}(j) \times BR \times w(j,i)
\end{equation}

where $dN^{p}/dydp_T$ is the total yield of the primary protons,
$dN^{m}/dydp_T$ -- the total yield of the measured protons,
$dN^\Lambda/dydp_T$ -- the total yield of the measured $\Lambda$,
BR -- branching ratio of the $\Lambda$ decay $\Lambda\rightarrow p\pi^-$,
$i$ is the $p_T$ bin number,
$N_{bins}$ is the number of bins,
$w(j,i)$ is the fraction of protons from $\Lambda$ decay from bin number $j$ which fall
into the proton bin number $i$. These fractions were extracted from MC.
The feed--down corrected proton and antiproton
$p_T$ spectra are shown in Fig.~\ref{fig:protons}.
We calculated the total yield, $dN/dy$, for protons and antiprotons 
by fitting them to the same  distribution as used for the $\Lambda$
and integrating from zero to infinity.
The results for the total yields $dN/dy$,
the average transverse momentum $<\!\!p_T\!\!>$
and the slope parameters $T$
for minimum--bias and for the 5$\%$ most central events
are also listed in Table~\ref{tab:table1}.
The measured $\Lambda/p$ and $\bar{\Lambda}/\bar{p}$ ratios after feed--down
corrections are found to be 0.89~$\pm$~0.07 (stat) and 0.95~$\pm$~0.09 (stat).

The net baryon numbers are indicative of the baryon transport mechanism in
relativistic heavy ion (RHI) collisions. In Table~\ref{tab:table2} we compare our
results for minimum--bias and for
the 5$\%$ most central events with the predictions of the
HIJING \cite{hijing} model
which assumes that the primary mechanism for baryon transport in RHI is due to
quark--diquark hadronic strings.
Vance et al. \cite{vance} implemented a non--perturbative
gluon--junction mechanism in a new version of HIJING (called HIJING/B)
to explain the enhanced baryon stopping at CERN SPS energies.
The predictions of this model are also shown in Table~\ref{tab:table2}.
For a valid comparison with the experimental data, the HIJING and
HIJING/B results for $\Lambda$ and $\bar{\Lambda}$ also include the
feed--down from the heavier hyperons, and the results for protons and
antiprotons the feed--down from $\Sigma^{\pm}$.
The Table shows a clear
difference (by a factor of $\sim$~4)
between HIJING and HIJING/B for the net $\Lambda$ number, with
the HIJING/B predictions in much better agreement, in particular  
for the minimum--bias data, where the experimental 
errors are relatively small.
The difference between the two models for the net proton number
is less obvious and although the data seem to favor HIJING/B the present 
accuracy does not allow for a clear preference between the two models.

In conclusion, we have measured the transverse momentum spectra of the
of $\Lambda$ and $\bar{\Lambda}$ particles in the $p_T$ range 0.4~$< p_T~<$~1.8~GeV/c
in minimum--bias and the 5$\%$ most central Au+Au collisions at
$\sqrt{s_{_{NN}}}~=~130$~GeV
at RHIC with the PHENIX experiment.
The absolute yields of $dN/dy$, at mid--rapidity, of $\Lambda$ hyperons are determined by
extrapolating to all values of $p_T$. The average
$\bar{\Lambda}/\Lambda$ ratio is found to be 0.75~$\pm$~0.09.
The ratio is constant over the whole $p_T$ range and there is also no 
significant variation of the $\bar{\Lambda}/\Lambda$ ratio 
as a function of the number of participants.
Using the measured $\Lambda$ and $\bar{\Lambda}$ yields, 
the $p$ and $\bar{p}$ yields corrected 
for feed--down from $\Lambda$ decays are determined.
The $\Lambda/p$ and $\bar{\Lambda}/\bar{p}$ ratios after feed--down
corrections are found to be 0.89~$\pm$~0.07 and 0.95~$\pm$~0.09.
The measured net $\Lambda$ number is substantially larger than
predicted by HIJING as already seen at CERN SPS and which may indicate 
enhanced baryon stopping at RHIC energies. 
The newly available high statistics data at $\sqrt{s_{_{NN}}}~=~200$~GeV
will allow us to further study baryon transport and strangeness
production in RHI collisions.

%\section{Acknowledgements}

We thank the staff of the Collider--Accelerator and Physics Departments at
BNL for their vital contributions.  We acknowledge support from the
Department of Energy and NSF (U.S.A.), MEXT and JSPS (Japan), RAS,
RMAE, and RMS (Russia), BMBF, DAAD, and AvH (Germany), VR and KAW
(Sweden), MIST and NSERC (Canada), CNPq and FAPESP (Brazil), IN2P3/CNRS
(France), DAE and DST (India), KRF and CHEP (Korea), the U.S. CRDF for 
the FSU, and the US--Israel BSF.

%\vspace{4cm}

%%%% FIGURE 1. Lambda and antilambda spectra
%\vspace{1cm}

%%%% FIGURE 2. Absolute yields
%\vspace{1cm}

%%%% FIGURE 3. Ratios
%\vspace{1cm}

%%%% FIGURE 4. Corrected protons
\vspace{10cm}

\vspace{10cm}
\begin{table}
\caption[]{ Inclusive $\Lambda$ and feed--down corrected proton yields, average transverse momentum (in GeV/c)
                  and slope parameters (in MeV/c)
                  for minimum--bias (MB) and for the 5$\%$ most central events.
                  The errors listed are statistical.
                  The systematic errors are 17$\%$ for the protons and 16$\%$ for the $\Lambda$.
                }
\begin{tabular}[ ]{cccccc}
&  &  &  \\
& $\Lambda$ & $\bar{\Lambda}$ & $p$ & $\bar{p}$ \\
&  &  &  \\
\hline
&  &  &  \\
$dN/dy$ (MB) & 4.8 $\pm$ 0.3 & 3.5 $\pm$ 0.3 & 5.4 $\pm$ 0.2 & 3.7 $\pm$ 0.2 \\
~~~~~~~~~~(5$\%$) & 17.3 $\pm$ 1.8 & 12.7 $\pm$ 1.8 & 19.3 $\pm$ 0.6 & 13.7 $\pm$ 0.7 \\
&  &  &  \\
$<\!\!p_T\!\!>$ (MB) & 1.06 $\pm$ 0.08 & 1.10 $\pm$ 0.12 & 0.88 $\pm$ 0.04 & 0.91 $\pm$ 0.06 \\~~~~~~~~~~(5$\%$) & 1.15 $\pm$ 0.15 & 1.14 $\pm$ 0.21 & 0.95 $\pm$ 0.07 & 1.04 $\pm$ 0.12 \\&  &  &  \\ 
$T$ ~~~~~~(MB) & 355 $\pm$ 11 & 366 $\pm$ 13 & 292 $\pm$ 21 & 304 $\pm$ 23 \\
~~~~~~~~~~(5$\%$) & 384 $\pm$ 16 & 380 $\pm$ 19 & 319 $\pm$ 31 & 327 $\pm$ 34 \\ 
&  &  &  \\
\end{tabular}
\label{tab:table1}
\end{table}

%\vspace{1cm}
\begin{table}
\caption[]{Total measured and predicted baryon yields for minimum--bias (MB)
and for the 5$\%$ most central events.
The errors listed are statistical. The proton yields are corrected for feed--down.
The systematic errors are 24$\%$ for the protons and 23$\%$ for the $\Lambda$.}
\begin{tabular}[ ]{cccc}
&  &  &  \\
Net baryon number & PHENIX & HIJING & HIJING/B \\
&  &  &  \\
\hline
&  &  &  \\
($\Lambda$ -- $\bar{\Lambda}$) ~~~~~(MB) & 1.3  $\pm$ 0.4 & 0.2 & 0.8 \\
&  &  &  \\
($p$ -- $\bar{p}$) ~~~~~(MB) & 1.7  $\pm$ 0.3 & 1.1 & 1.7 \\
&  &  &  \\
($\Lambda$ -- $\bar{\Lambda}$) ~~~~~(5$\%$) & 4.6  $\pm$ 2.5 & 0.8 & 3.2 \\
&  &  &  \\
($p$ -- $\bar{p}$) ~~~~~~(5$\%$) & 5.6  $\pm$ 0.9 & 4.7 & 7.1 \\
&  &  &  \\
\end{tabular}
\label{tab:table2}
\end{table}

\end{multicols}


\begin{references}
\vspace{-1.6cm}

\bibitem[*]{Deceased}Deceased     % for Leo Paffrath and Sten Garpman

\bibitem[\dagger]{non-par}Not a participating Institution.

\bibitem{sqm-2000} 5$^{th}$ International Conference on Strangeness in Quark Matter 2000,
J. Phys. G {\bf 27}, 255 (2001) and references therein.

\bibitem{kopeliovich} B.Z. Kopeliovich and B.G. Zakharov, Z. Phys. C {\bf 43}, 241 (1989).

\bibitem{kharzeev} D. Kharzeev, Phys. Lett. B {\bf 378}, 238 (1996).

\bibitem{vance} S.E. Vance {\it et al.}, Phys. Lett. B {\bf 443}, 45 (1998).

\bibitem{gyalassy} M. Gyulassy and I. Vitev, nucl-th/0104066.

\bibitem{ahmed} S.Ahmed {\it et al.}, Phys. Lett. B {\bf 382}, 35, (1996).

\bibitem{filimonov} K. Filimonov {\it et al}, Nucl. Phys. {\bf A661}, 166c (1999).

\bibitem{na35} T. Alber {\it et al.}, Eur. J. Phys. {\bf C2}, 643 (1998).

\bibitem{na44} K. Wolf {\it et al.}, Phys. Rev. C {\bf 57}, 837 (1998).

\bibitem{na49} H. Appelsh{\"a}user {\it et al.}, Phys. Rev. Lett. {\bf 82}, 2471 (1999).

\bibitem{phenix_1} PHENIX Collaboration, D.P. Morrison {\it et al.}, 
Nucl. Phys. {\bf A638}, 565c (1998).

\bibitem{pc} P. Nilsson {\it et al.}, Nucl. Phys. {\bf A661}, 665c (1999).

\bibitem{emcal} PHENIX Collaboration, S. White {\it et al.}, Nucl. Phys. 
{\bf A698}, (2002).

\bibitem{ppg001} K. Adcox {\it et al.}, Phys. Rev. Lett. {\bf 86}, 3500 (2001).

\bibitem{ppg003} K. Adcox {\it et al.}, Phys. Rev. Lett. {\bf 88}, 022301 (2002).

\bibitem{geant} GEANT 3.21, CERN program library.

\bibitem{thermal} P. Braun-Munzinger {\it et al.}, 
Phys. Lett. B {\bf 518}, 41 (2001).

\bibitem{ppg006} K. Adcox {\it et al.}, nucl-ex/0112006.

\bibitem{hijing} X.N. Wang and M. Gyulassy, Phys. Rev. D {\bf 44}, 3501 
(1991).

\end{references}
\end{document}